\documentclass[numberedappendix,apj,twocolumn,letter]{emulateapj}

\usepackage{amssymb}
\usepackage{amsmath}
\newcommand{\be}{\begin{equation}}
\newcommand{\ee}{\end{equation}}

\shorttitle{Connecting GRB formation to galaxy properties} 
\shortauthors{Trenti et al.}

\begin{document}


\title{Gamma Ray Burst and star formation rates: The physical origin
  for the redshift evolution of their ratio}

\author{Michele Trenti\altaffilmark{1,\dag},
Rosalba Perna\altaffilmark{2}, Sandro Tacchella\altaffilmark{3}}

\altaffiltext{1}{Institute of Astronomy and Kavli Institute for Cosmology, University of  Cambridge, Madingley Road, Cambridge,  CB3 0HA, United Kingdom}
\altaffiltext{2}{JILA, Department of Astrophysical and Planetary Sciences, University of Colorado, 389-UCB, Boulder, CO 80309 USA}
\altaffiltext{3}{Department of Physics, Institute for Astronomy, ETH Zurich, CH-8093 Zurich, Switzerland}
\altaffiltext{\dag}{Kavli Institute Fellow}
\email{trenti@ast.cam.ac.uk} 

%

\begin{abstract}
  
  Gamma Ray Bursts (GRBs) and galaxies at high redshift represent
  complementary probes of the star formation history of the Universe.
  In fact, both the GRB rate and the galaxy luminosity density are
  connected to the underlying star formation.  Here, we combine a star
  formation model for the evolution of the galaxy luminosity function
  from $z=0$ to $z=10$ with a metallicity-dependent efficiency for GRB
  formation to simultaneously predict the comoving GRB rate. Our model
  sheds light on the physical origin of the empirical relation often
  assumed between GRB rate and luminosity density-derived star
  formation rate:
  $\dot{n}_{GRB}(z)=\varepsilon(z)\times\dot{\rho}^*_{obs}(z)$, with
  $\varepsilon(z)\propto(1+z)^{1.2}$. At $z\lesssim4$,
  $\varepsilon(z)$ is dominated by the effects of metallicity
  evolution in the GRB efficiency.  Our best-fitting model only
  requires a moderate preference for low-metallicity, that is a GRB
  rate per unit stellar mass about four times higher for
  $\log{(Z/Z_{\odot})}<-3$ compared to $\log{(Z/Z_{\odot})}>0$. Models with
  total suppression of GRB formation at $\log{(Z/Z_{\odot})}\gtrsim 0$
  are disfavoured. At $z\gtrsim4$, most of the star formation happens
  in low-metallicity hosts with nearly saturated efficiency of GRB
  production per unit stellar mass. However at the same epoch, galaxy
  surveys miss an increasing fraction of the predicted luminosity
  density because of flux limits, driving an accelerated evolution of
  $\varepsilon(z)$ compared to the empirical power-law fit from lower
  $z$. Our findings are consistent with the non-detections of GRB
  hosts in ultradeep imaging at $z>5$, and point toward current galaxy
  surveys at $z>8$ only observing the top $15-20\%$ of the total
  luminosity density.

\end{abstract}


\section{Introduction}\label{sec:intro}

The star formation history of the Universe is a fundamental observable
to understand the assembly and evolution of
galaxies, the production of ionizing radiation and the gas chemical
enrichment. Among the different (indirect) tracers of the star
formation rate, two of particular relevance at high redshift are the
measurement of the rest-frame UV luminosity density from galaxy
surveys (e.g.~\citealt{madau98,hopkins06,bouwens11}), and the rate of
long-duration gamma-ray bursts (GRBs)\citep{kistler09}.

Deep surveys with ground and space telescopes identified thousands of
galaxies up to redshift $z\lesssim 10$, when the Universe was just
about $500$ Myr old
\citep{shimasaku06,bouwens07,bouwens11,ouchi10,trenti11,bradley12,mclure13}. However,
converting luminosity density into a SFR depends upon stellar
population properties, such as metallicity \citep{madau98}. In
addition, dust obscuration is a severe problem for rest-frame U
observations, especially at $z\lesssim4$: The intrinsic luminosity
density might be up to ten times larger than the observed one and is
highly sensitive to dust correction estimates
\citep{smit12}. Moreover, surveys are flux limited, hence the derived
luminosity density traces the star formation rate (SFR) only in
galaxies above detection threshold ($\dot{\rho}^*_{obs}(z)$). Because
the galaxy luminosity function (LF) evolves with redshift, the amount
of missed star formation changes as well, likely becoming more severe
at high $z$, and is difficult to quantify directly
\citep{trenti10,robertson13}.

Long-duration GRBs ($t>2$s) are instead detectable at cosmological
distances \citep{tanvir09} and trace the SFR as well, since they are
generally associated to the collapse of massive stars
\citep{MacFadyen99}. However, the number of events is much
smaller compared to that of high-$z$ galaxies, and more importantly
there is likely a metallicity bias, related to a GRB production
mechanism dependent on progenitor properties
\citep{fruchter99,berger03,savaglio09,perley13}, for example because
of reduced mass-loss rates in massive, low-metallicity stars
\citep{yoon06}. The presence of such bias is supported by
observations of a majority of low-metallicity hosts (e.g.,
\citealt{savaglio09,graham12,jimenez13}), although multiple cases of
super-solar GRB hosts exist \citep{levesque10b,savaglio12,perley13}. A
further complication arises from understanding biases that dust
obscuration (metallicity and redshift dependent) may introduce into
the GRB rate ($\dot{n}_{GRB}(z)$) inference because of dark bursts
\citep{perley09,greiner11}.

Given the complementary strenghts and weakenesses of both approaches,
past investigations tried to connect them by typically assuming a
relation:
\begin{equation}\label{eq:GRB_rate}
\dot{n}_{GRB}(z)=\varepsilon(z)\times\dot{\rho}^*_{obs}(z),
\end{equation}
where $\varepsilon(z)$ is the efficiency of GRB production per unit
stellar mass, empirically assumed to have form:
\begin{equation}\label{eq:varepsilon}
\varepsilon(z)=\varepsilon_0 (1+z)^{\beta},
\end{equation}
with $\beta \approx 1.2$ calibrated at $z\lesssim4$ and then
extrapolated at higher redshift
\citep{kistler09,virgili11,robertson12}. The challenges with this
empirical modeling are twofolds. First, the GRB rate depends
physically on the \emph{total} SFR ($\dot{\rho}^*_{tot}(z)$), not on
$\dot{\rho}^*_{obs}(z)$ (derived from flux-limited galaxy surveys)
as used in Equation~\ref{eq:GRB_rate}, because the GRB afterglow is
much brighter than its host and thus can be detected independently of
host luminosity. Hence, one should have
$\dot{n}_{GRB}(z)\propto~\dot{\rho}^*_{tot}(z)$. Second, without a
physical model to link GRB and star formation, there is no guarantee
that Equation~\ref{eq:varepsilon} holds beyond the redshift range of
its calibration, raising concerns on systematic biases at $z\gtrsim6$.

Here, we investigate the physical origin of the connection between GRB
and SFR through $\varepsilon(z)$, and how this depends on metallicity
of progenitors and on the difference between $\dot{\rho}^*_{tot}(z)$
and $\dot{\rho}^*_{obs}(z)$. We resort to a simple, yet successful
model for the evolution of the galaxy UV LF (developed in
\citealt{tacchella13}; TTC13 hereafter), which we complement with the
mass-metallicity relation for host galaxies from \citet{maiolino08}
and with a metallicity-dependent GRB efficiency, inspired by stellar
evolution simulations \citep{yoon06}. The setup of our modeling is
described in Section~\ref{sec:model}, results in
Section~\ref{sec:results}, conclusions in Section~\ref{sec:con}. As in
TTC13, we adopt a WMAP5 cosmology: $\Omega_{\Lambda,0}=0.72$,
$\Omega_{m,0}=0.28$, $\Omega_{b,0}=0.0462$, $\sigma_8=0.817$,
$n_s=0.96$, $h=0.7$ \citep{komatsu09}.

\section{Star formation model and metal-dependent GRB formation}\label{sec:model}

The base for our study is the model presented in TTC13, which links
star formation to dark-matter halo assembly through a mass-dependent
efficiency on a timescale defined by the halo assembly time, adopting
a continuous star formation history over this timescale. Following
\citet{lacey93}, this is the time needed to grow the main progenitor
from halo mass $M_h/2$ to $M_h$. The key model feature is that the
efficiency of star formation, $\xi(M_h)=M_*/M_h$, where $M_*$ is the
stellar mass, depends only on halo mass and is redshift independent
(see also~\citealt{behroozi13}). Since the halo assembly time is
redshift dependent, the model naturally accounts for the redshift
evolution of the rest-frame UV luminosity, computed using the Single
Stellar Population models of \citet{bruzual03}, at fixed halo mass
(assembly times become shorter at high-$z$, thus halos are brighter).
The model also includes treatment of dust extinction, which affects
significantly the observed rest-frame UV luminosity of high-$z$
galaxies, implemented with an empirically calibrated formula following
\citet{smit12} (for comprehensive discussion see Sec.~3.3 in TTC13).

Given $\xi(M_h)$ and the DM halo mass function (from
\citealt{sheth99}), the model thus fully describes the evolution of
both the intrinsic galaxy LF, as well as that of the observed one
(with dust extinction). All details on the efficiency of converting
baryons into stars are encapsulated into $\xi(M_h)$. To calibrate this
relation, we resort, like in TTC13, to abundance matching at $z=4$:
Assuming that each DM halo hosts a single galaxy, we derive $\xi(M_h)$
so that the model, dust-extinct, LF has a Schechter form
($\phi(L)=\phi^*(L/L^*)^{\alpha}\exp{-(L/L^*)}$) with parameters:
$\phi^*=1.3\times10^{-3}~\mathrm{Mpc^{-3}}$; $\alpha=-1.73$ and
$M_{AB}^*=-21.0$, where $M^*=-2.5\log_{10}{L^*}$ \citep{bouwens07}.
Note that TTC13 account for scatter in the luminosity to halo-mass
relation, and for a burst of star formation at the halo assembly time
as additional free-parameter to calibrate. Both features only add
modest improvements on the data-model comparison for the LF and star
formation rates, so we neglect them here. Our simplified model is
still capable of an overall good description of the star formation
rate evolution over the whole history of the Universe
(Fig.~\ref{fig:rates}, left panel), and allows us to explore the link
to the production of GRB events while minimizing the free parameters.

Compared to other recent investigations on the connection between GRB
production and SFR \citep{robertson12,hao13}, our framework has the
advantage of relying on a physical model for the galaxy LF, therefore
allowing us to explore the effects of star formation below the
observational limit of the deepest galaxies survey. This way, we can
address what is the contribution to the GRB production by dwarf-like
galaxies at high-$z$ which are too faint for direct observations, with
high complementarity to the recent modeling by \citet{jimenez13} at
$z\lesssim3$ by means of reconstruction of the star formation history
from SDSS observations of local galaxies. 

In addition, to investigate whether and how much GRB production is
preferentially located in low-metallicity environments, we augment our
model using the mass-metallicity relation of \citet{maiolino08} to
assign galaxy metallicities. For this, we adopt their Equation~2
with coefficients from Table~5 (for \citealt{bruzual03} models) and
linear interpolation in redshift space. Then, we resort to stellar
evolution simulations by \citet{yoon06} to construct a basic form for
the metallicity-dependent efficiency of GRB formation. The main
limitation is that, because simulations of GRB progenitors are
computationally expensive, \citet{yoon06} only explore a limited
number of models, with coarse sampling of metallicity (see their
Figs.~3~and~6). Broadly, they conclude that very low metallicity
progenitors ($Z\lesssim10^{-3}~Z_{\odot}$) are two-three times more
likely to produce GRBs compared to low metallicity ones
($Z\sim10^{-1}~Z_{\odot}$), and that efficiency drops to $\sim0$
for higher content of metals. The latter conclusion is however in
tension with recent observations of GRB hosts with $Z\sim~Z_{\odot}$
\citep{levesque10b,elliott13}, and might be related to the limited
parameter space explored by the \citet{yoon06}
simulations. Alternatively, channels for GRB production which are
different from the Collapsar model and do not require low metallicity
progenitors have been proposed, such as binary systems
\citep{fryer05,podsiadlowski10}. Therefore, we include in our
metallicity-dependent efficiency a plateau of non-zero probability of
forming GRBs which is metallicity-independent, treating this as a free
parameter to explore by comparison with the observed GRB rate. The
piece-wise linear functional form for our metal-dependent efficiency
of GRB production is:
\begin{equation}\label{eq:metal_eff}
  \kappa(Z)=\kappa_0\times\frac{a\log_{10}{Z/Z_{\odot}}+b+p}{1+p},
\end{equation}
where $\kappa_0$ and $p$ are free parameters (normalization of the
relation and efficiency plateau at $Z\geq~Z_{\odot}$), while $a$ and
$b$ depend on $Z$ as follows. For $Z/Z_{\odot}\leq10^{-3}$, $a=0$,
$b=1$; for $10^{-3}\leq Z/Z_{\odot}\leq10^{-1}$, $a=-3/8$, $b=-1/8$;
for $10^{-1}\leq Z/Z_{\odot}\leq1$, $a=-1/4$, $b=0$; for
$Z/Z_{\odot}>1$, $a=0$, $b=0$. Fig.~\ref{fig:metallicity_dep} shows
$\kappa(Z)$ for $\kappa_0=1$ and $p=0.3$ (our fiducial plateau value).

For a model with given threshold in minimum halo mass for star
formation and given efficiency of forming GRBs in high-metallicity
environments (parameter $p$), we determine $\kappa_0$ (normalization
of GRB efficiency production) by carrying out a least-square fit of
our predicted GRB rate by comparison with \citet{wanderman10}
(Fig.~\ref{fig:rates}, right panel).  We limit the fit to points at
$z\leq6$ because of growing uncertainty in observations at higher
redshift.

\section{Modeling result}\label{sec:results}

Our canonical model assumes that star formation proceeds down to halos
with mass $M_h\geq 10^8~\mathrm{M_{\odot}}$, approximately equivalent
to virial temperatures $T_{vir}\gtrsim10^4~\mathrm{K}$, sufficient for
cooling and forming stars even in presence of a UV background, albeit
at very low efficiency \citep{trenti09,finlator11}. We predict that
these halos host very faint galaxies with magnitude
$-12\lesssim~M_{UV}\lesssim-10$, depending on redshift.\footnote{For
  comparison, the deepest observations in the HUDF field reach
  $M_{UV}\sim-17$ \citep{bouwens07}, but recently \citet{alavi13} used
  gravitational lensing to probe the $z\sim2$ UV LF down to
  $M_{UV}\sim-13$, demonstrating that it remains a steep power law,
  like in our model.}  For the efficiency of GRB production with
metallicity, we assume $p=0.3$ in Equation~\ref{eq:metal_eff}, so that
we form a (small) fraction of GRBs in high-metallicity environments.
The results of the comparison of the model predictions to the data are
shown in Fig.~\ref{fig:rates}. The model (black-dashed line) captures
well the evolution of the observed SFR $\dot{\rho}^*_{obs}(z)$ (data
points from a variety of surveys, see Fig.~2 in TTC13 for full
details), derived from the luminosity density for galaxies with
$M_{AB}\leq-17.7$. At $z\lesssim3$, integrating over the model LF to
$M_{AB}\leq-11$ (thus including faint galaxies), yields an
approximately constant increase of the SFR by a factor $\sim1.3$
(red-solid line). At higher redshift there is a growing fraction of
star formation in small mass, low luminosity halos, so that at
$z\gtrsim8$ only $\lesssim 20\%$ of all star formation is seen in
current surveys (see also \citealt{trenti10} for similar
conclusions). The model prediction for the GRB rate, which derives
from the SFR for all galaxies, weighted by the metallicity preference
for low-$Z$ hosts, is shown as red-solid line in the right panel of
Fig.~\ref{fig:rates} and describes the data well.

In Fig.~\ref{fig:psi} we use these results to investigate our
predictions for the relation $\varepsilon(z)$ between the SFR and GRB
rate (Equation~\ref{eq:GRB_rate}), empirically modeled as power law in
redshift by past studies (Equation~\ref{eq:varepsilon}).  Our model
(red-solid line) predicts indeed an approximate power law with
$\beta=1.2$, as derived by \citet{virgili11,robertson12}, in the
redshift range where the large majority of GRBs are observed
($1\lesssim z\lesssim5$).  We predict deviations from a power law both
in the local Universe and at very high $z$, with our model staying
above $(1+z)^{1.2}$. To understand the physical origin of
$\varepsilon(z)$, we decompose the contribution to this quantity in
metallicity effect (green-dashed line) and faint galaxies (blue-dashed
line). We see immediately that metallicity evolution is the main
driver of $\varepsilon(z)$ at $z\lesssim5$. At higher $z$, though,
most of the star formation happens in low-luminosity, small mass
galaxies, where the mass-metallicity relation predicts low $Z$ and
near-maximal GRB production per unit stellar mass. Therefore, based on
metallicity alone one would predict a flattening of $\varepsilon(z)$
at high $z$. However, around the same redshift at which the
metal-dependent efficiency of GRB production saturates, the fraction
of missed star formation in faint galaxies begins to increase
steadily. This is the driver of the accelerated growth of
$\varepsilon(z)$ at $z\gtrsim5$. Finally, we note that at low redshift
our model predicts a slight excess of GRBs compared to the observed
GRB rate (Fig.~\ref{fig:rates}) and to
$\varepsilon(z)\sim(1+z)^{\beta}$. While we stress that there is no
physical reason because $\varepsilon(z)$ should be a power law, it is
nevertheless possible that a more accurate parameterization of
$\kappa(Z)$ would give better agreement.

Overall, our canonical model provides a comprehensive description of
the data both for $\dot{\rho}^*_{obs}(z)$ and for
$\dot{n}_{GRB}(z)$. To further understand how unique/robust these
predictions are, it is useful to compare them to those obtained by
varying model parameters. This is illustrated for $\dot{n}_{GRB}(z)$
in Fig.~\ref{fig:alt_models}, for a range of different assumptions
that all give by construction the same description of the
\emph{observed} SFR $\dot{\rho}^*_{obs}(z)$.
Fig.~\ref{fig:alt_models}, left panel, explores the effects of varying
the metallicity-dependence in GRB production. The blue line shows
predictions for complete suppression of GRBs in high metallicity
progenitors ($p=0$), while the green line has no metallicity
dependence ($p\to+\infty$). It is immediate to see that both scenarios
represent a worse description of the data ($\chi^2=9.5$ and
$\chi^2=5.6$ respectively) compared to the canonical model (red,
$\chi^2=2.1$), with systematic deviations which introduce
autocorrelation in the residuals at low and high redshift (going into
opposite directions). The right panel illustrates the same three
models, but here we assume that there is no star formation in galaxies
with $M_{AB}>-17.7$ (this means
$\dot{\rho}^*_{obs}(z)\equiv\dot{\rho}^*_{total}(z)$). Scenarios with
both extreme and no metallicity dependence are similarly disfavored
($\chi^2=383$ and $\chi^2=6.9$ respectively). Interestingly the
current data from $\dot{n}_{GRB}(z)$ alone don't rule out that GRBs
originate from star formation sites with $M_{AB}\lesssim-17.7$
($\chi^2=1.9$). In fact, at $z\lesssim5$ the models with and without
faint galaxies are essentially yielding the same predictions, just
with a different $\kappa_0$ normalization.  However, as we discussed
in \citet{trenti12}, searches for host-galaxies of GRBs at high-$z$
are a powerful probe to investigate whether star formation is
happening in low-luminosity galaxies. From the non-detection of $n=6$
GRB host galaxies at $z>5$, we can confidently exclude this
alternative scenario (\citealt{trenti12}; see also
\citealt{tanvir12}). We thus conclude that by combining all available
observations, the best model is the one predicting a growing abundance
of faint galaxies as the redshift increases, along with a moderate
preference for GRBs to form in low-metallicity environments
($p\approx0.3$). Without duplicating the discussion of
\citet{trenti10,trenti12}, this strengthens the conclusion that the
majority of the ionizing photons at $z>6$ are produced by galaxies
currently too faint to be observed in ultradeep fields with the Hubble
Space Telescope (see also \citealt{shull12}).

\section{Conclusions}\label{sec:con}

In this \emph{letter} we investigated the relation between star
formation and GRB rate. For this, we resorted to a physical model
which describes the galaxy LF evolution from $z=0$ to $z=10$,
augmented it with a mass-metallicity relation, and assumed a
metallicity-dependent efficiency for GRB production. We showed that we
can construct a successful GRB-rate model assuming that star formation
continues down to galaxies with $M_{AB}\lesssim-11$ and that GRBs are
about four times more likely to originate in very low metallicity
environments ($Z/Z_{\odot}<10^{-3}$) compared to
$Z\sim~Z_{\odot}$. With this model we investigated the physical origin
for the redshift evolution of the ratio between GRB and SFR
($\varepsilon(z)$, Equation~\ref{eq:varepsilon}). We showed that the
approximate $(1+z)^{1.2}$ behavior empirically derived by previous
studies at $z\lesssim 4$ is driven primarily by metallicity
dependence.  Importantly, our model predicts a steepening of such
relation at higher $z$, because of a growing fraction of star
formation happening in galaxies with $M_{AB}>-17.7$ during the epoch
of reionization at $z\gtrsim5$.

\begin{figure}
\begin{center} 
\includegraphics[scale=0.39]{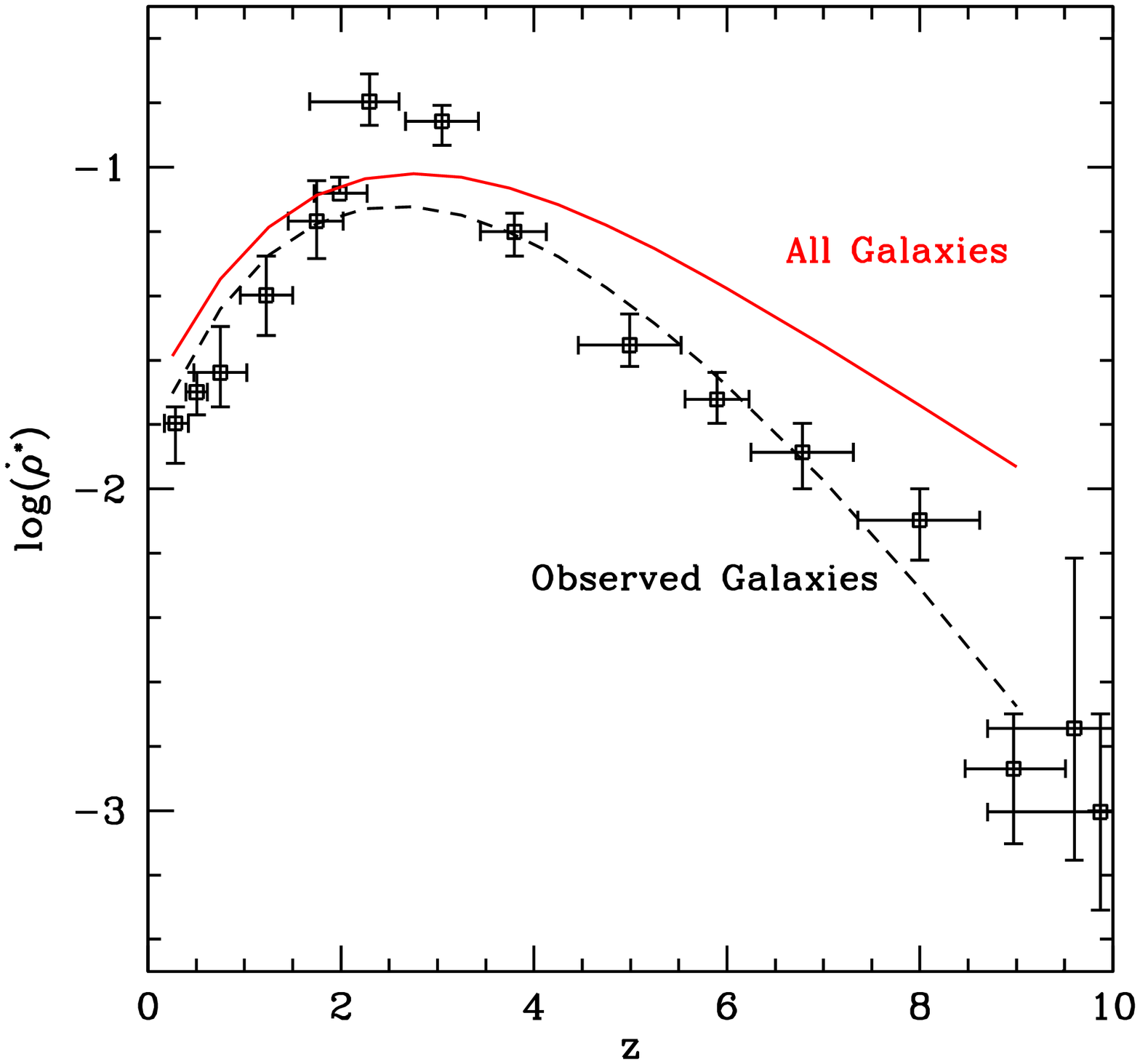}
\includegraphics[scale=0.39]{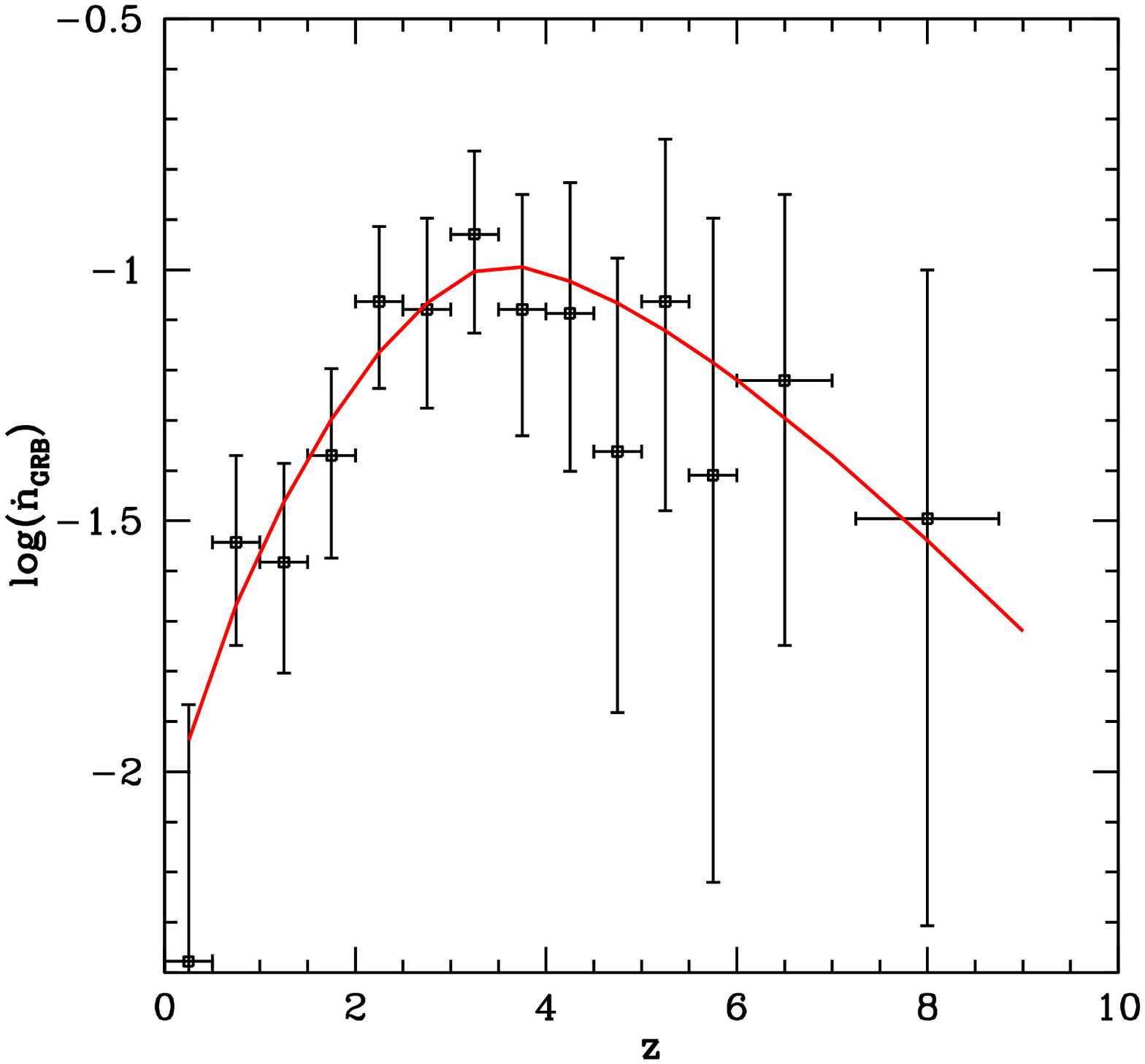}
\end{center}
\caption{Left panel: Star formation rate versus redshift, inferred
  from LBG observations in the UV integrated to $M_{AB}=-17.7$ and
  corrected for dust extinction (see TTC13). The black-dashed line
  shows predictions from our LF model integrated to the same
  observational limit. The red-solid line shows instead the model with
  LF integrated over all galaxies ($M_{AB}\lesssim-11$). Right panel:
  GRB comoving rate from \citet{wanderman10} compared to predictions
  of our best fitting model, which includes a moderate
  metallicity-dependence on the GRB rate (see
  Fig.~\ref{fig:metallicity_dep}).  }\label{fig:rates}
\end{figure}

\begin{figure}
\begin{center} 
\includegraphics[scale=0.4]{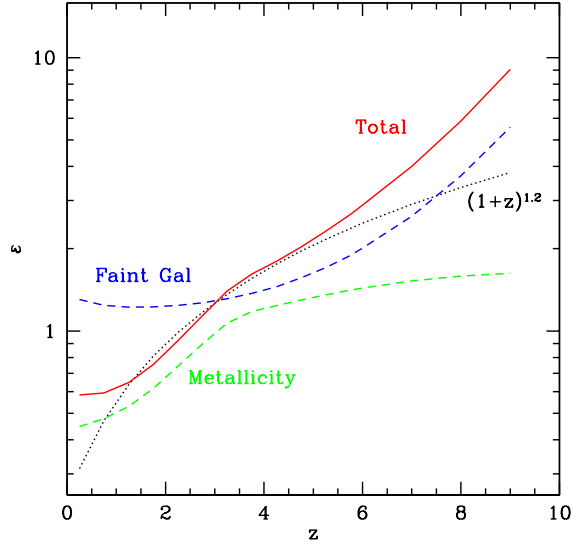}
\end{center}
\caption{Redshift evolution of the GRB rate to SFR ratio
  $\varepsilon(z)$ (red-solid line). Our model highlights that at low
  redshift the metallicity evolution is the main driver of
  $\varepsilon(z)$ (green-dashed line), while at high-$z$ changes in
  $\varepsilon(z)$ are primarily due to an increasing fraction of
  integrated light missed in flux-limited LBG surveys (blue-dashed
  line). At $z>5$ we see an accelerated evolution of $\varepsilon(z)$
  compared to the best fitting power-law $(1+z)^{1.2}$ derived at
  $z\lesssim5$ (black-dotted line).}\label{fig:psi}
\end{figure}

\begin{figure}
\begin{center} 
\includegraphics[scale=0.39]{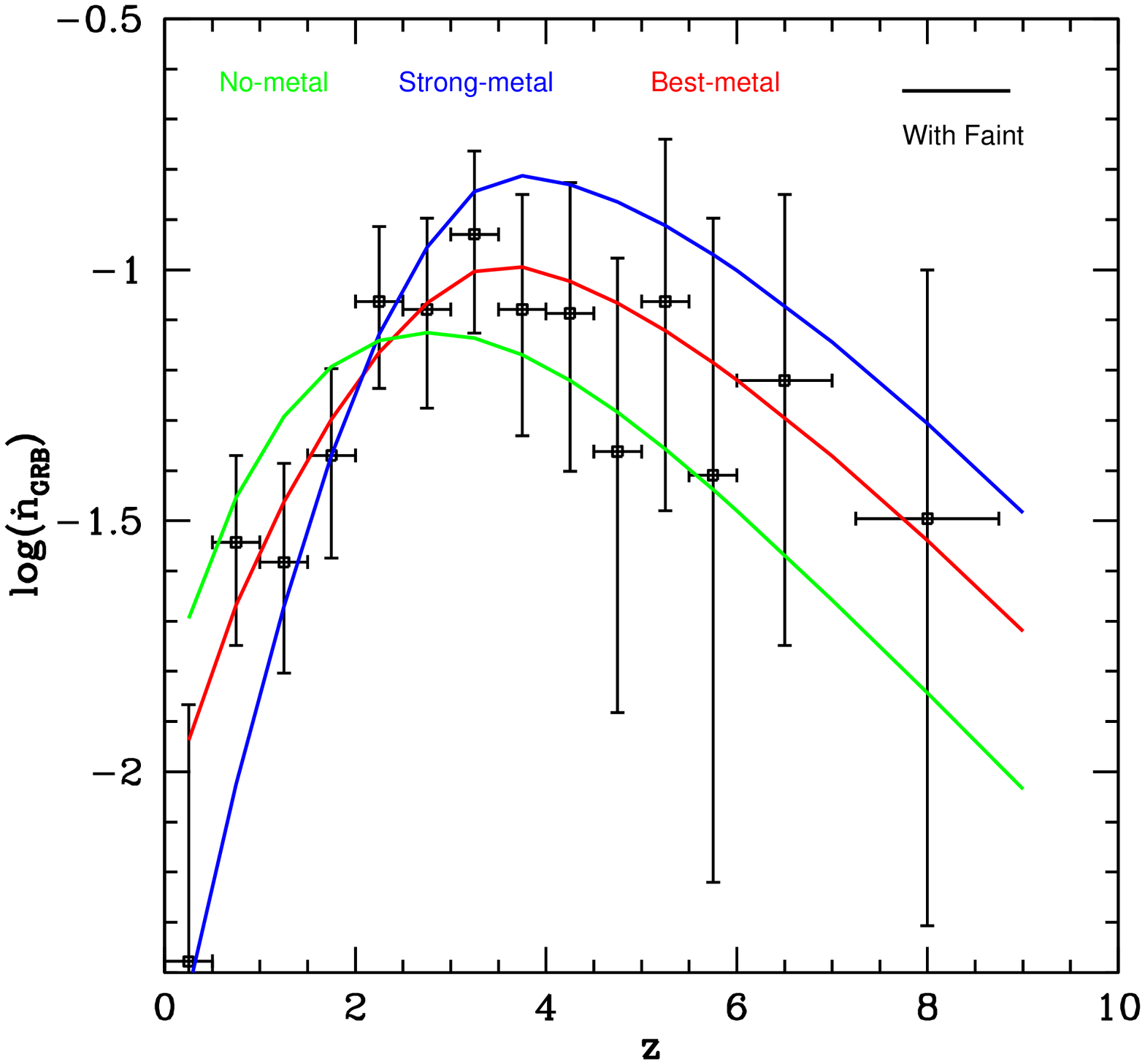}
\includegraphics[scale=0.39]{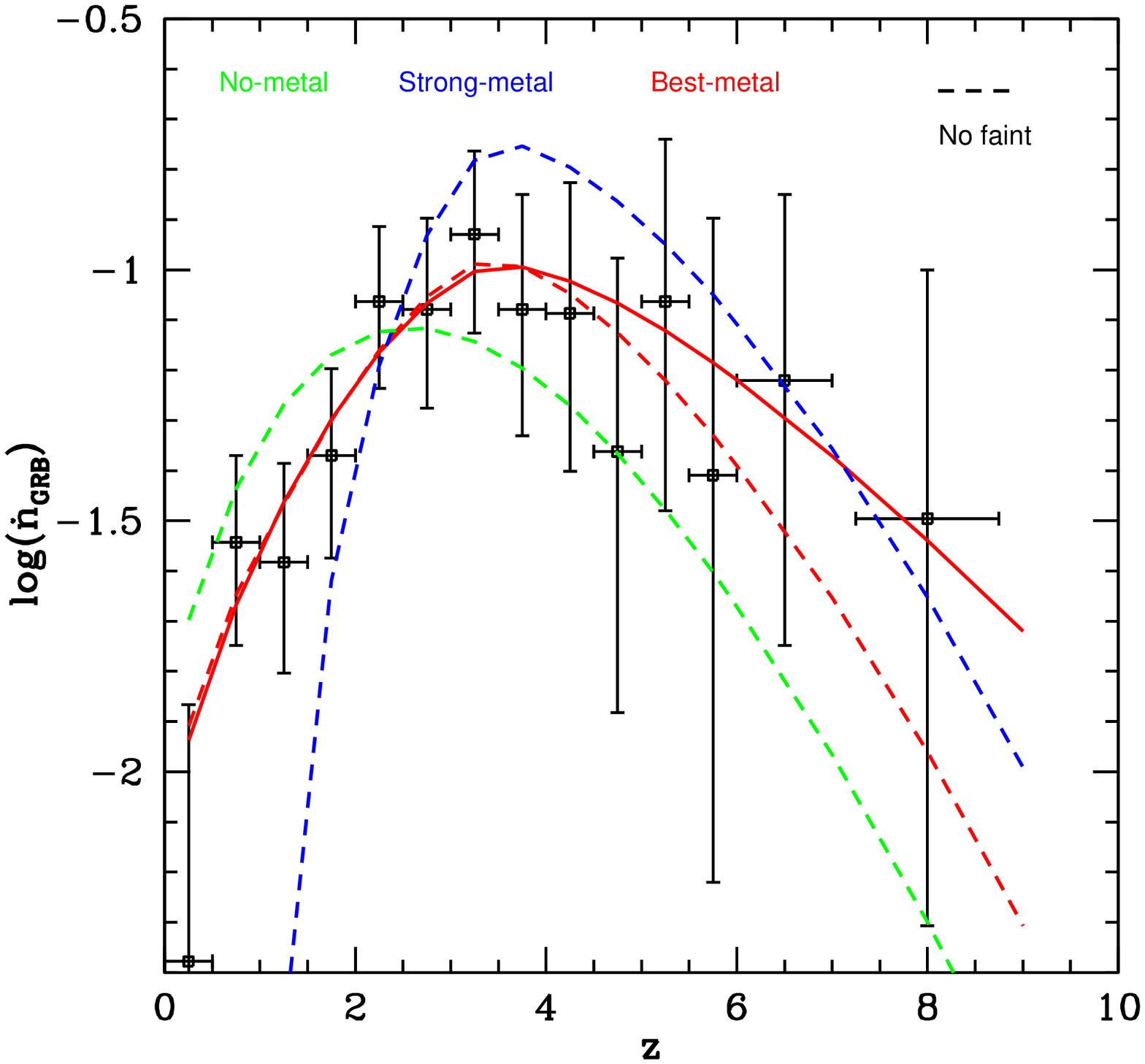}
\end{center}
\caption{Left panel: GRB comoving rate for models with different
  efficiencies for forming GRBs as function of metallicity, compared
  to the \citet{wanderman10} rate. The red solid line is our reference
  model; the blue line represents a model with complete suppression of
  GRB formation in $Z>Z_{\sun}$ environments, and the green line is a
  model without metallicity dependence ($\kappa(Z)$ constant). Right
  panel: same as left panel but for models where star formation
  happens only in galaxies with $M_{AB}\leq-17.7$.  A model with
  moderate metallicity dependence still provides an acceptable fit
  without assuming star formation below the observational limit
  (red-dashed line). Such model is however in strong tension (ruled
  out at $>99$\% confidence) with the non-detection of GRB host
  galaxies at $z>5$ (see
  \citealt{trenti12,tanvir12}).}\label{fig:alt_models}
\end{figure}

\begin{figure}
\begin{center} 
\includegraphics[scale=0.4]{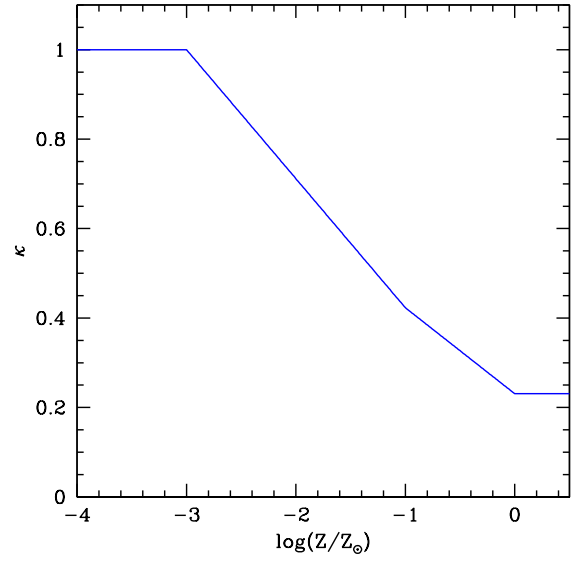}
\end{center}
\caption{Functional form for efficiency of GRB production per unit
  stellar mass versus host-galaxy metallicity for our fiducial
  model. GRBs are about four times more likely in low metallicity
  progenitors compared to super-solar ones.}\label{fig:metallicity_dep}
\end{figure}

Our model can be used to derive an improved estimate of the total SFR
at $z>4$ from GRB events, compared to past investigations which rely
on extrapolation of $\varepsilon(z)\propto(1+z)^{\beta}$ beyond $z>4$
(Fig.~\ref{fig:rates}). In this respect, we note that we are obtaining
a self-consistent description of both SFR and GRB rate, as well as of
the stellar mass density with redshift (the latter is discussed in
TTC13, e.g. see their Fig.~2). In contrast, \citet{robertson12} argue
that the GRB-inferred star formation rate is inconsistent with the
stellar mass assembly. Their result is likely due to the extrapolation
for $\varepsilon(z)$, which leads to an overestimate of the SFR
inferred from GRBs (since their $\varepsilon(z)$ is underestimated at
$z>4$ based on our conclusions). Overall, we can conclude that there
is no tension between using either the GRB rate or the observed
luminosity density from galaxy surveys as tracers of star formation,
provided that the presence of faint galaxies below the current
detection limits is taken into account.

Of course, the uncertainties on the data, especially on the GRB rate,
are still large. Thus, it is fundamental to identify additional
independent tests of the framework we introduced. For example,
\citet{jimenez13} assume that GRB production has a hard cut-off at
$Z/Z_{\odot}\geq0.1$ and obtain a good match to
$\dot{n}_{GRB}(z\lesssim3)$ considering the star formation history at
$z\lesssim3$ reconstructed from stellar archeology observations of
galaxies in the local Universe. At higher redshift ($z\gtrsim3$), some
GRBs might have Population III progenitors forming in rare pockets of
metal free gas, predicted theoretically \citep{trenti09}, and recently
observed \citep{fumagalli11}. To better discriminate among all these
different models, a promising approach would be to use our model to
construct predictions for how the LF of GRB hosts at $z\sim 1-3$
differs from the one observed for Lyman-break galaxies. With a growing
sample of GRB hosts being identified by nearly-complete follow-up
surveys such as TOUGH \citep{hjorth12} such comparison should be
achievable in the near future. This will provide validation/refinement
for our current treatment of the metallicity bias. In addition,
ultradeep follow-up observations of host galaxies of GRBs at
$z\gtrsim5$ have the potential to quantify the fraction of missed star
formation in galaxy surveys such as the Hubble Ultradeep Field
\citep{bouwens11}, providing a highly complementary tool to
investigate star formation during the epoch of reionization without
waiting for next generation facilities such as the \emph{James Webb
  Space Telescope}.

\acknowledgements 

We thank Raul Jimenez, Roberto Maiolino, Max Pettini, and Sandra
Savaglio for useful discussions. This work was partially supported by
the European Commission through the Marie Curie Career Integration
Fellowship PCIG12-GA-2012-333749 (MT) and by NSF Grant No. AST 1009396
(RP).





\end{document}